\documentclass[10pt]{article}
\usepackage{color}
\usepackage{epsfig}
\usepackage{graphicx}
\usepackage{subfigure}
\usepackage{ulem}
\usepackage{amsfonts}
\usepackage{epsfig,bm}
\usepackage{graphicx}
\usepackage{comment}
\newcommand{\be}{\begin{equation}}
\newcommand{\ee}{\end{equation}}
\newcommand{\ba}{\begin{eqnarray}}
\newcommand{\ea}{\end{eqnarray}}
\newcommand{\bs}{\begin{scriptsize}}
\newcommand{\es}{\end{scriptsize}}
\newcommand{\nn}{\nonumber\\}
\begin{document}
\begin{center}
\title[ \textbf{State-space geometry, non-extremal black holes and Kaluza-Klein
monopoles}\\
\vspace{0.5cm} {{\bf  Stefano Bellucci\footnote{e-mail:
bellucci@lnf.infn.it }$^{,a}$ and Bhupendra Nath
Tiwari\footnote{e-mail: tiwari@lnf.infn.it}$^{,a}$}}
\end{center}
\vspace{0.3cm}
\begin{center}
{$^a$ \it INFN- Laboratori Nazionali di Frascati,\\
 Via E. Fermi 40, 00044, Frascati, Italy}\\
\end{center}
\vspace{0.3cm} We examine the statistical nature of the charged
anticharged non-extremal black holes in string theory. From the
perspective of the intrinsic Riemannian Geometry, the first
principle of the statistical mechanics shows that  the stability
properties of general nonextremal nonlarge charged black brane
solutions are divulged from the positivity of the corresponding
principle minors of the space-state metric tensor. Under the
addition of the Kaluza-Klein monopoles, a novel aspect of the
Gaussian fluctuations demonstrates that the canonical fluctuations
can be ascertained without any approximation. We offer the
state-space geometric implication for the most general
non-extremal black brane configurations in string theory.


Generic higher charged non-extremal black branes in string theory
\cite{9601029v2,9411187v3,9504147v2,0409148v2,9707203v1,0507014v1,0502157v4,
0505122v2} and $M$-theory \cite{0209114,0401129,0408106,0408122}
possess rich state-space geometric structures. Some examples of
such state-space configurations involve statistical properties of
the extremal and non-extremal black branes
\cite{0606084v1,0801.4087v1,SST,BNTBull,BNTBullcorr,BNTBull08,RotBH,BSBR}.
In this paper, we focus our attention on the thermodynamic
perspectives of the higher charged anticharged black brane
configurations in string theory. We wish to explicate the nature
of the state-space pair correlation functions and the associated
stability properties of the higher charged black brane solutions
containing an ensemble of branes and antibranes. In the past,
there have been several notions analyzed in condensed matter
physics \cite{RuppeinerRMP,RuppeinerA20,RuppeinerPRL,RuppeinerA27,
RuppeinerA41}. Here, we shall consider eight charged anticharged
string theory black brane configurations and analyze the
state-space pair correlation functions and their relative scaling
relations. Given the definite state-space description of
consistent non-extremal black brane macroscopic solutions, we
expose (i) for what conditions the considered black hole
configuration is stable, (ii) how its state-space correlation
functions scale in terms of the numbers of the branes and
antibranes. In sequel, we enlist the complete set of non-trivial
relative state-space correlation functions of the nonextremal
nonlarge charged anticharged black brane configurations. See for
an introduction references
\cite{0801.4087v1,SST,BNTBull,BNTBull08,RotBH,BSBR}. A similar
analysis remains valid for the black holes in general relativity
\cite{gr-qc/0601119v1, gr-qc/0512035v1, gr-qc/0304015v1,
0510139v3}, attractor black holes
\cite{9508072v3,9602111v3,new1,new2,
0702019v1,0805.1310,BFGM2,Kallosh-rev,bfm6} and Legendre
transformed finite parameter chemical configurations
\cite{Weinhold1, Weinhold2}, quantum field theory and hot QCD
backgrounds \cite{BNTSBVC,bntsbvc1} with finite chemical
potentials and the strongly coupled quarkonium configurations
\cite{bntsbvc2, bntsbvc3}.

Before analyzing the state-space properties of the eight parameter
black brane configuration, let us first provide a brief
introduction to the thermodynamic geometry \cite{RuppeinerRMP,
0510139v3, 0606084v1}. From the perspective of the intrinsic
geometry, the state-space geometry is defined as the thermodynamic
geometry with a set of non-equilibrium coordinates, viz. the
number of the branes and antibranes. In this framework, the
state-space metric tensor is defined as the negative Hessian
matrix of the black hole entropy $S(\vec{x})$. In general, the
components of the metric tensor are defined as
\begin{eqnarray}
g_{ij}:=-\frac{\partial^2 S(\vec{x})}{\partial x^j \partial x^i}.
\end{eqnarray}
In the above definition, a state-space covariant vector $\vec{x}
\in M_n $ shall be understood as the collection of the charges and
anticharges $(n_i, m_i)$ of the considered black hole. We shall
show that the state-space geometry thus defined takes an account
of the local thermodynamic interactions and possible global vacuum
phase transitions. To illustrate the consideration of state-space
geometry, let us explicate the case of two parameter black brane
configurations. To do so, let the two parameters be the charge $n$
and the anticharge $m$, then the components of the Ruppeiner
metric tensor are
\begin{eqnarray}
g_{nn}= - \frac{\partial^2 S}{\partial n^2},\ \ 
g_{nm}= - \frac{\partial^2 S}{{\partial n}{\partial m}},\ \ 
g_{mm}= - \frac{\partial^2 S}{\partial m^2}.
\end{eqnarray}
In this se-up, the components of the state-space metric tensor are
associated to the respective statistical pair correlation
functions. It is worth mentioning that the co-ordinates on the
state-space manifold are the parameters of the microscopic
boundary conformal field theory which is dual the black hole
space-time solution. This is because the underlying state-space
metric tensor comprises of the Gaussian fluctuations of the
entropy which is the function of the number of the branes and
antibranes. For the chosen black hole configuration, the local
stability of the underlying statistical system requires both
principle minors to be positive. In this case, the diagonal
components of the state-space metric tensor, viz., $\{ g_{x_ix_i}
\mid x_i= (n,m)\}$ signify the heat capacities of the system. This
requires that the diagonal components of the state-space metric
tensor
\begin{eqnarray}
g_{x_ix_i} &>& 0, \ i= \ n, m
\end{eqnarray}
be positive definite. In this investigation, we discuss the
significance of the above observation for the eight parameter
non-extremal black brane configurations in string theory. From the
notion of the relative scaling property, we demonstrate the nature
of the brane-brane pair correlations. From the perspective of the
intrinsic Riemannian geometry, the stability properties of the
eight parameter black branes are examined from the positivity of
the principle minors of the space-state metric tensor. For the
Gaussian fluctuations of the two charge equilibrium statistical
configurations, the existence of a positive definite volume form
on the state-space manifold $(M_2(R),g)$ imposes such a global
stability condition. In particular, the above configuration leads
to a stable statistical basis, if the determinant of the
state-space metric tensor
\begin{eqnarray}
\Vert g \Vert &= &S_{nn}S_{mm}- S_{nm}^2
\end{eqnarray}
remains positive. For the two charge black brane configurations,
the geometric quantities corresponding to the underlying
state-space manifold elucidates typical features of the Gaussian
fluctuations about an ensemble of equilibrium brane microstates.
Subsequently, we can further calculate the Christoffel connection
$\Gamma_{ijk}$, Riemann curvature tensor $R_{ijkl}$, Ricci tensor
$R_{ij}$ and Ricci scalar $ R $ for the intrinsic state-space
manifold. From the above viewpoint, the intrinsic scalar curvature
accompanies information of the global correlation volume of the
underlying statistical systems. In the case of the two charge
black hole configurations, we have the following scalar curvature
\begin{eqnarray} R= \frac{1}{2 \Vert g \Vert^2}
(S_{mm}S_{nnn}S_{nmm}+ S_{nm}S_{nnm}S_{nmm}+ S_{nn}S_{nnm}S_{mmm}
-S_{nm}S_{nnn}S_{mmm}- S_{nn}S_{nmm}^2- S_{mm}S_{nnm}^2 ).
\end{eqnarray}
In this picture, the zero scalar curvature signifies that the
informations on the event horizon of the black hole fluctuate
independently of each other, while a divergent scalar curvature
indicates vacuum phase transitions. This leads to the fact that an
ensemble of highly correlated pixels of information can have
vacuum phase transitions on the event horizon of the black hole.
Ruppeiner has further interpreted the assumption ``that all the
statistical degrees of freedom of a black hole live on the black
hole event horizon'' as indicating that the state-space scalar
curvature signifies the average number of correlated Planck areas
on the event horizon of the black hole \cite{RuppeinerRMP}. From
the perspective of Mathur's fuzzball proposal \cite{0502050v1},
the present consideration takes an account of the fact that the
area of the event horizon is an integral multiple of the Planck
area \cite{Bekenstein}.

Interestingly, notice further that the state-space scalar
curvature explicates the nature of the long range global phase
transitions. In this sense, an ensemble of microstates corresponds
to the black hole states, which are statistically (i) interacting,
if the underlying state-space configuration has a non-zero scalar
curvature and (ii) non-interacting, if the  scalar curvature
vanishes identically. Incrementally, one may note that the
state-space configuration of the eight charge-anticharge black
hole is attractive or repulsive, and weakly interacting in
general. Subsequently, we shall demonstrate that the above  eight
charge-anticharge black hole yield a stable statistical
configuration, if at most three of the parameters, viz., the brane
numbers $\{n_i\}$, and the antibrane numbers $\{m_i\}$ are allowed
to fluctuate.

With this introduction of the two dimensional intrinsic
state-space geometry, we shall proceed to systematically analyze
the underlying statistical structures of the higher charged black
hole configurations in string theory. Following the notions we
have developed in the Refs.
\cite{0801.4087v1,SST,BNTBull,BNTBull08,BNTBullcorr,RotBH,BSBR},
the subsequent analysis is devoted to the state-space geometric
implications of the eight charge non-extremal black branes. In the
present work, we examine the nature of the state-space of the
non-extremal black hole under the contribution of finitely many
non-trivially circularly fibered KK-monopoles. In this process, we
shall also enlist the complete set of non-trivial relative
state-space correlation functions of the configurations considered
in \cite{BNTBull,BNTBull08}. In the past, there have been
calculations of the entropy of the extremal, near-extremal and
general nonextremal solutions \cite{9603195v1, 9603061v2} in
string theory. Inductively, the most general charge anticharge
nonextremal black hole has the following entropy

\ba S(n_1,m_1,n_2,m_2,n_3,m_3,n_4,m_4) &=& 2 \pi
\prod_{i=1}^4(\sqrt{n_i}+\sqrt{m_i}) 
\ea

For the distinct $i,i,k \in \{1,2,3,4\}$, we find that the
components of the metric tensor are

\ba g_{n_in_i} &=& \frac{\pi}{2n_i^{3/2}} \prod_{j \ne i}
(\sqrt{n_j}+\sqrt{m_j}), \nn
g_{n_in_j} &=& -\frac{\pi}{2 (n_i n_j)^{1/2}} \prod_{k \ne i \ne
j} (\sqrt{n_k}+\sqrt{m_k}), \nn
g_{n_im_i} &=& 0, \nn
g_{n_im_j} &=& -\frac{\pi}{2 (n_i m_j)^{1/2}} \prod_{k \ne i \ne
j} (\sqrt{n_k}+\sqrt{m_k}), \nn
g_{m_im_i} &=& \frac{\pi}{2 m_i^{3/2}} \prod_{j \ne i}
(\sqrt{n_j}+\sqrt{m_j}),\nn
g_{m_im_j} &=& -\frac{\pi}{2 (m_i m_j)^{1/2}} \prod_{k \ne i \ne
j} (\sqrt{n_k}+\sqrt{m_k}).
\ea
From the above depiction, it is evident that the principle
components of the state-space metric tensor $\lbrace g_{n_in_i},
g_{m_im_i} \vert \ i=1,2,3,4 \rbrace$ essentially signify a set of
definite heat capacities (or the related compressibilities) whose
positivity in turn apprises that the black brane solutions comply
with an underlying equilibrium statistical configuration. For an
arbitrary number of the branes $\{n_i\}$ and antibranes $\{m_i\}$,
we find that the associated state-space metric constraints as the
diagonal pair correlation functions remain positive definite.  In
particular, $\forall \ i \in \{1,2,3,4 \}$, it is clear that we
have the following positivity conditions
\begin{eqnarray}
g_{n_in_i}> 0 \ \mid n_i, m_i > 0, \ \ 
g_{m_im_i}> 0 \ \mid n_i, m_i > 0.
\end{eqnarray}
As observed in \cite{BNTBull,BNTBull08}, we find that the ratios
of diagonal components vary inversely with a multiple of a
well-defined factor in the underlying parameters, viz., the
charges and anticharges, which change under the Gaussian
fluctuations, whereas the ratios involving off diagonal components
in effect uniquely inversely vary in the parameters of the chosen
set $A_i$ of equilibrium black brane configurations. This suggests
that the diagonal components weaken in a relatively controlled
fashion into an equilibrium, in contrast with the off diagonal
components which vary over the domain of associated parameters
defining the $D_1$-$D_5$-$P$-$KK$ non-extremal non-large charge
configurations. In short, we can easily substantiate for the
distinct $x_i:=(n_i, m_i) \mid i \in \lbrace 1,2,3,4 \rbrace $
describing eight (anti)charge string theory black holes that the
relative pair correlation functions have distinct types of
relative correlation functions. Apart from the zeros, infinities
and similar factorizations, we see that the non-trivial relative
correlation functions satisfy the following scaling relations
\begin{eqnarray}
\frac{g_{x_ix_i}}{g_{x_jx_j}}&=& (\frac{x_j}{x_i})^{3/2}, \nonumber \\
\frac{g_{x_ix_j}}{g_{x_kx_l}}&=& (\frac{x_ix_j}{x_kx_l})^{-1/2}
(\frac{\prod_{p \ne i \ne j} \sqrt{n_p}+\sqrt{m_p}}{\prod_{q \ne
k \ne l} \sqrt{n_q}+\sqrt{m_q}}), \nonumber \\
\frac{g_{x_ix_i}}{g_{x_ix_k}}&=&-\sqrt{(\frac{x_k}{x_i})}
(\frac{\prod_{p \ne i }\sqrt{n_p}+\sqrt{m_p}}{\prod_{q \ne i \ne
k} \sqrt{n_q}+\sqrt{m_q}}).
\end{eqnarray}
%
As noticed in the Refs. \cite{BNTBull,BNTBull08}, it is not
difficult to analyze the statistical stability properties of the
eight charged string theory non-extremal black holes. In
particular, one can easily determine the principle minors
associated with the state-space metric tensor and thereby argue
that all the principle minors must be positive definite, in order
to have a globally stable configuration. In the present case, it
turns out that the above black hole is stable only when some of
the charges and/or anticharges are held fixed or take specific
values such that $p_i > 0$ for all the dimensions of the
state-space manifold. Subsequently, from the definition of the
Hessian matrix of the associated entropy concerning the most
general nonextremal nonlarge charged black holes, we observe that
some of the principle minors $p_i$ are indeed non-positive. In
fact, we discover an uniform local stability criteria on the lower
dimensional hyper-surfaces and two dimensional surface of the
underlying state-space manifold. The corresponding principle
minors take the following explicit expressions
\ba p_1 &=& \frac{\pi}{2 n_1^{3/2}} (\sqrt{n_2}+\sqrt{m_2})
(\sqrt{n_3}+\sqrt{m_3}) (\sqrt{n_4}+\sqrt{m_4}), \nn
p_2&=& \frac{\pi^2}{4(n_1 m_1)^{3/2}} (\sqrt{n_2}+\sqrt{m_2})^2
(\sqrt{n_3}+\sqrt{m_3})^2 (\sqrt{n_4}+\sqrt{m_4})^2, \nn
p_3&=& \frac{\pi^3}{8(n_1 m_1 n_2)^{1/2}}
(\sqrt{n_3}+\sqrt{m_3})^3 (\sqrt{n_4}+\sqrt{m_4})^3
(\sqrt{n_2}+\sqrt{m_2}) (\sqrt{n_1}+\sqrt{m_1}), \nn
p_4&=&0, \nn
p_5&=& -\frac{\pi^5}{8(n_1 n_2 m_2 m_1)^{3/2} n_3}
(\sqrt{n_2}+\sqrt{m_2}) (\sqrt{n_3}+\sqrt{m_3})^3
(\sqrt{n_4}+\sqrt{m_4})^5 (\sqrt{n_1}+\sqrt{m_1})^2 \tilde{p}_5
,\nn
p_6&=& -\frac{\pi^6}{16 (n_1 n_2 m_1 m_2 n_3 m_3)^{3/2}}
(\sqrt{n_2}+\sqrt{m_2})^2 (\sqrt{n_3}+\sqrt{m_3})^3
(\sqrt{n_4}+\sqrt{m_4})^6 (\sqrt{n_1}+\sqrt{m_1})^3 \tilde{p}_6
,\nn
p_7&=& -\frac{\pi^7}{32 (n_1 m_1 n_2 m_2 n_3 m_3 n_4)^{3/2}}
(\sqrt{n_2}+\sqrt{m_2})^3 (\sqrt{n_3}+\sqrt{m_3})^3
(\sqrt{n_4}+\sqrt{m_4})^5 (\sqrt{n_1}+\sqrt{m_1})^4  \tilde{p}_7,
\nn
p_8&=& -\frac{\pi^8}{16 (\prod_{i=1}^4 n_i m_i)^{3/2}}
(\sqrt{n_2}+\sqrt{m_2})^4 (\sqrt{n_3}+\sqrt{m_3})^4
(\sqrt{n_4}+\sqrt{m_4})^5 (\sqrt{n_1}+\sqrt{m_1})^5  \tilde{p}_8.
\ea
The exact expression of the factors $\{\tilde{p}_i \}$ of the
higher principle minors, viz., $\{p_i \ \vert \ i=5,6,7,8\}$ is
relegated to the Appendix.
\begin{figure*}
\includegraphics[height=.20\textheight,angle=0]{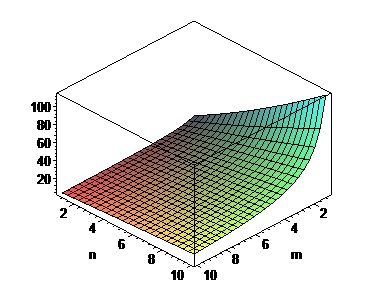} \hspace{1cm}
\includegraphics[height=.20\textheight,angle=0]{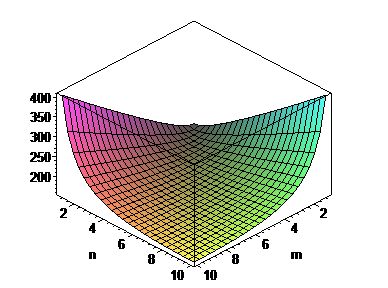}
\caption{The heat capacities as the diagonal components $g_{ii}$
of the state-space metric tensor and the surface minor $p_2$,
plotted as the functions of the number of branes $n$ and
antibranes $m$, describing the fluctuations in the statistical
configuration.}  \label{corrloc1}
\end{figure*}
\begin{figure*}
\includegraphics[height=.20\textheight,angle=0]{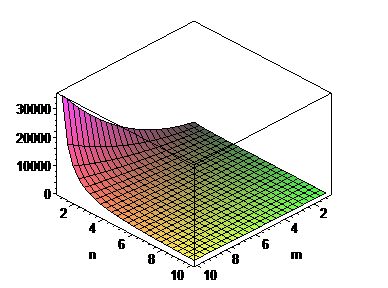} \hspace{1cm}
\includegraphics[height=.20\textheight,angle=0]{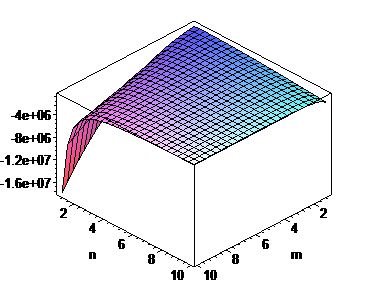}
\caption{The hypersurface minors $p_3$ and $p_5$ of the
state-space metric tensor, plotted as the functions of the number
of branes $n$ and antibranes $m$, describing the fluctuations in
the statistical configuration.} \label{corrloc2}
\end{figure*}
\begin{figure*}
\includegraphics[height=.20\textheight,angle=0]{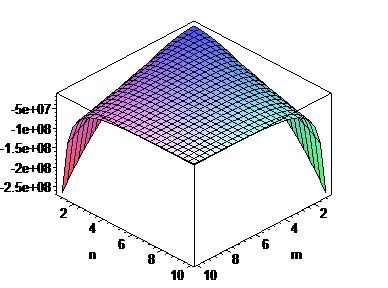} \hspace{1cm}
\includegraphics[height=.20\textheight,angle=0]{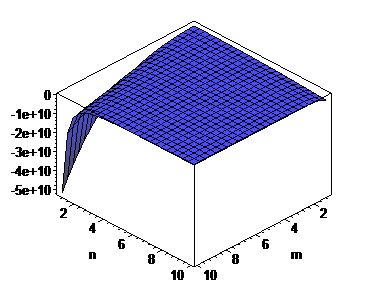}
\caption{The hypersurface minors $p_6$ and $p_7$ of the
state-space metric tensor, plotted as the functions of the number
of branes $n$ and antibranes $m$, describing the fluctuations in
the statistical configuration.} \label{corrloc3}
\end{figure*}
\begin{figure*}
\includegraphics[height=.20\textheight,angle=0]{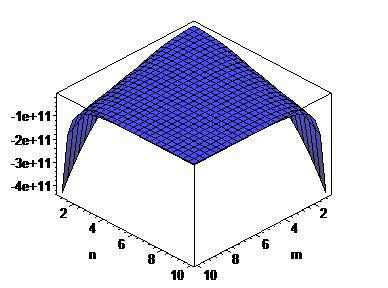} \hspace{1cm}
\includegraphics[height=.20\textheight,angle=0]{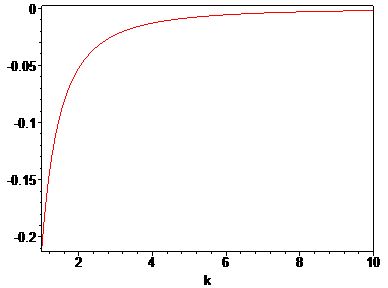}
\caption{The determinant as the highest principle minor $p_8$ of
the state-space metric tensor plotted as the function of the
number of branes $n$ and antibranes $m$, and the corresponding
scalar curvature for the equal number of branes and antibranes
$(n= m)$ describing the fluctuations in the statistical
configuration.} \label{corrglob}
\end{figure*}
As per the above evaluation, the graphical perspective of the
state-space quantities is offered for the $n$ branes and $m$
antibranes. In fact, we have obtained the exact expressions for
the components of the metric tensor, principle minors, determinant
of the metric tensor and the underlying scalar curvature of the
fluctuating statistical configuration of the eight parameter black
holes in string theory. Qualitatively, the local and the global
correlation properties of the limiting vacuum configuration are
shown in Figs.(\ref{corrloc1}, \ref{corrloc2}, \ref{corrloc3},
\ref{corrglob}). Under the statistical fluctuations, the first
three figures, \textit{viz.}, Figs.(\ref{corrloc1},
\ref{corrloc2}, \ref{corrloc3}), describe the local stability
properties, and the first of the last figure, \textit{viz.},
Fig.(\ref{corrglob}), describes the global ensemble stability,
whereas the second one describes the corresponding phase space
stability of the eight parameter black hole configuration.

In general, there exists an akin higher degree polynomial equation
on which the Ricci scalar curvature becomes null, and exactly on
these points the state-space configuration of the underlying
non-large charge nonextremal eight charge black hole system
corresponds to a non-interacting statistical system. Here, the
state-space manifold $(M_8,g)$ is curvature free. A systematic
calculation further shows that the general expression for the
Ricci scalar is quite involved, and even for equal brane charges
$n_1:=n$; $n_2:=n$; $n_3:=n$;  $n_4:=n$ and equal antibrane
charges $m_1:=m$; $m_2:=m$; $m_3:=m$, $m_4:=m$ the result does not
sufficiently simplify. Nevertheless, we find for the identical
large values of brane and antibrane charges $n:=k$ and $m:=k$
\cite{BNTBull, BNTBull08} that there exists an attractive
state-space configuration. In the limit of a large $k$, the
corresponding scalar curvature reduces to the following small
negative value \ba R = -\frac{21}{32} \frac{1}{\pi k^2} \ea

Interestingly, it turns out that the system becomes noninteracting
in the limit of $k \rightarrow \infty$. For the case of the $n= k=
m$, we observe that the corresponding principle minors reduce to
the following constant values
\ba \{p_i\}_{i=1}^8 &=& \{4 \pi, 16 \pi^2, 32 \pi^3, 0, -2048
\pi^5, -16384 \pi^6, -163840 \pi^7, -1048576 \pi^8 \}.\ea
In this case, we find that the limiting underlying statistical
system remains stable when at most three of the parameters,
\textit{viz.}, $\{n_i=n= m_i\}$, are allowed to fluctuate.
Herewith, we find that the state-space manifold of the eight
parameter brane and antibrane configuration is free from critical
phenomena, except for the roots of the determinant. Thus, the
regular state-space scalar curvature is comprehensively universal
for the non-large charge non-extremal black brane configurations
in string theory. In fact, the above perception turns out to be
justified from the typical state-space geometry, \textit{viz.},
the definition of the metric tensor as the negative Hessian matrix
of the duality invariant expression of the black brane entropy. In
this case, we may nevertheless easily observe, for a given entropy
$S_0$, that the constant entropy curve is given by the following
curve
\begin{eqnarray}
(\sqrt{n_1} + \sqrt{m_1}) (\sqrt{ n_2 }+ \sqrt{m_2}) (\sqrt{n_3}+
\sqrt{m_3}) (\sqrt{ n_4 }+ \sqrt{m_4}) = c.
\end{eqnarray}
where $ c $ is a real constant taking precisely the value
$S_0/2\pi$. Under the vacuum fluctuations, the present analysis
indicates that the entropy of the eight parameter black brane
solution defines a non-degenerate embedding in the viewpoints of
intrinsic state-space geometry. The above computations further
encourage that our state-space geometry determines an intricate
set of statistical properties, \textit{viz.}, pair correlation
functions and correlation volume, which reveal the possible nature
of the associated parameters prescribing an ensemble of
microstates of the dual CFT living on the boundary of the black
brane solution. Furthermore, our expectation is that we can
consider such an analysis for all higher dimensional black brane
configurations with multiple parameters, where the state-space
geometric propositions of having ordinary computations might be a
bit involved. However, one may exhibit the intrinsic geometric
acquisitions with an appropriate comprehension of the required
parameters defining the state-space coordinate for the chosen
black brane configurations in the string theory.

We have analyzed state-space pair correlation functions and the
notion of stability for the non-extremal black holes in string
theory. Our consideration is from the viewpoints of intrinsic
state-space geometry. From the intrinsic Riemannian geometry, we
find that the stability of these black branes has been divulged
from the positivity of principle minors of the space-state metric
tensor. Following developments introduced in the Refs.
\cite{0801.4087v1,SST,BNTBull,BNTBull08,BNTBullcorr,RotBH,BSBR},
we have explicitly extended the analysis of the state-space
geometry for the four charge and four anticharge non-extremal
black brane configurations in the string theory. The present
consideration of the eight parameter black brane configurations,
where the underlying leading order statistical entropy is written
as a function of the charges $ \lbrace n_i \rbrace $ and
anticharges $ \lbrace m_i \rbrace $ and describes the stability
properties under the Gaussian fluctuations. The present
consideration includes all the special cases of the extremal and
near-extremal configurations with a fewer number of charges and
anticharges. In this case, we obtain the standard pattern of the
underlying state-space geometry and constant entropy curve as that
of the lower parameter non-extremal black holes. In fact, the
conclusion to be drawn remains the same, as the underlying
state-space geometry remains well-defined as an intrinsic
Riemannian manifold $ N:= M_8 \setminus \tilde{B}$, where $
\tilde{B} $ is the set of roots of the determinant of the metric
tensor. The local coordinate of the state-space manifold involves
the four charges and four anticharges of the underlying
non-extremal black holes. Our analysis indicates that the leading
order statistical behavior of the black brane configurations in
string theory remains intact under the inclusion of the
KK-monopoles.

\section*{Acknowledgement}

This work has been supported in part by the European Research
Council grant n.~226455, \textit{``SUPERSYMMETRY, QUANTUM GRAVITY
AND GAUGE FIELDS (SUPERFIELDS)"}.

This work was conducted during the period B.N.T. served as a
postdoctoral research fellow at the \textit{``INFN-Laboratori
Nazionali di Frascati, Roma, Italy''}.

\section*{Appendix}
In this appendix, we provide explicit forms of the higher
principle minors of the state-space metric tensor of the eight
charged nonextremal nonlarge black holes. Our analysis illustrates
that the stability properties of the specific state-space
hypersurface may exactly be exploited in general. The definite
behavior of state-space properties, as accounted in the concerned
main section suggests that the various intriguing hypersurfaces of
the state-space configuration include the nice feature that they
do have definite stability properties, except for some specific
values of the charges and anticharges. As mentioned in the main
sections, these configurations are generically well-defined and
indicate an interacting statistical basis. Herewith, we discover
that the state-space geometry of the general black brane
configurations in string theory indicate the possible nature of
the underlying statistical fluctuations. Significantly, we notice
from the very definition of the intrinsic metric tensor that the
related factors of the principle minors take the following
expressions

\begin{eqnarray}
\tilde{p}_5&:=& n_2 \sqrt{m_1}+2 \sqrt{n_2} \sqrt{m_2}
\sqrt{m_1}+m_2 \sqrt{m_1}+\sqrt{n_1} n_2+2 \sqrt{n_1} \sqrt{n_2}
\sqrt{m_2}+\sqrt{n_1} m_2,\nn
\tilde{p}_6&:=& n_2 \sqrt{m_1} \sqrt{n_3}+n_2 \sqrt{m_1}
\sqrt{m_3}+2 \sqrt{n_2} \sqrt{m_1} \sqrt{m_2} \sqrt{n_3}+2
\sqrt{n_2} \sqrt{m_1} \sqrt{m_2} \sqrt{m_3} \nn && +\sqrt{m_1} m_2
\sqrt{n_3}+\sqrt{m_1} m_2 \sqrt{m_3}+\sqrt{n_1} n_2
\sqrt{n_3}+\sqrt{n_1} n_2 \sqrt{m_3} \nn && +2 \sqrt{n_1}
\sqrt{n_2} \sqrt{m_2} \sqrt{n_3}+2 \sqrt{n_1} \sqrt{n_2}
\sqrt{m_2} \sqrt{m_3}+\sqrt{n_1} m_2 \sqrt{n_3}+\sqrt{n_1} m_2
\sqrt{m_3},\nn
\tilde{p}_7&:=& 2 \sqrt{n_1} n_2 \sqrt{n_3} \sqrt{m_3}
\sqrt{m_4}+2 m_2 \sqrt{n_3} \sqrt{m_3} \sqrt{m_4} \sqrt{m_1}+8
\sqrt{n_2} \sqrt{m_2} n_3 \sqrt{n_4} \sqrt{m_1} \nn && +8
\sqrt{n_1} \sqrt{n_2} \sqrt{m_2} n_3 \sqrt{n_4}+8 m_2 \sqrt{n_3}
\sqrt{m_3} \sqrt{n_4} \sqrt{m_1}+16 \sqrt{n_2} \sqrt{m_2}
\sqrt{n_3} \sqrt{m_3} \sqrt{n_4} \sqrt{m_1} \nn && +2 \sqrt{n_2}
\sqrt{m_2} n_3 \sqrt{m_4} \sqrt{m_1}+2 \sqrt{n_1} \sqrt{n_2}
\sqrt{m_2} m_3 \sqrt{m_4}+4 \sqrt{n_1} \sqrt{n_2} \sqrt{m_2}
\sqrt{n_3} \sqrt{m_3} \sqrt{m_4} \nn && +8 n_2 \sqrt{n_3}
\sqrt{m_3} \sqrt{n_4} \sqrt{m_1}+8 \sqrt{n_1} n_2 \sqrt{n_3}
\sqrt{m_3} \sqrt{n_4}+8 \sqrt{n_1} m_2 \sqrt{n_3} \sqrt{m_3}
\sqrt{n_4} \nn && +8 \sqrt{n_1} \sqrt{n_2} \sqrt{m_2} m_3
\sqrt{n_4}+8 \sqrt{n_2} \sqrt{m_2} m_3 \sqrt{n_4} \sqrt{m_1}+2
\sqrt{n_2} \sqrt{m_2} m_3 \sqrt{m_4} \sqrt{m_1} \nn && +2
\sqrt{n_1} \sqrt{n_2} \sqrt{m_2} n_3 \sqrt{m_4}+16 \sqrt{n_1}
\sqrt{n_2} \sqrt{m_2} \sqrt{n_3} \sqrt{m_3} \sqrt{n_4}+2
\sqrt{n_1} m_2 \sqrt{n_3} \sqrt{m_3} \sqrt{m_4} \nn && +\sqrt{n_1}
n_2 n_3 \sqrt{m_4}+4 \sqrt{n_1} m_2 m_3 \sqrt{n_4}+n_2 m_3
\sqrt{m_4} \sqrt{m_1}+4 m_2 n_3 \sqrt{n_4} \sqrt{m_1} \nn && +4
\sqrt{n_1} m_2 n_3 \sqrt{n_4}+\sqrt{n_1} m_2 m_3 \sqrt{m_4}+4 m_2
m_3 \sqrt{n_4} \sqrt{m_1}+4 n_2 m_3 \sqrt{n_4} \sqrt{m_1} \nn &&
+m_2 n_3 \sqrt{m_4} \sqrt{m_1}+m_2 m_3 \sqrt{m_4}
\sqrt{m_1}+\sqrt{n_1} n_2 m_3 \sqrt{m_4}+4 \sqrt{n_1} n_2 m_3
\sqrt{n_4} \nn && +4 \sqrt{n_1} n_2 n_3 \sqrt{n_4}+\sqrt{n_1} m_2
n_3 \sqrt{m_4}+4 n_2 n_3 \sqrt{n_4} \sqrt{m_1}+n_2 n_3 \sqrt{m_4}
\sqrt{m_1} \nn && +2 n_2 \sqrt{n_3} \sqrt{m_3} \sqrt{m_4}
\sqrt{m_1}+4 \sqrt{n_2} \sqrt{m_2} \sqrt{n_3} \sqrt{m_3}
\sqrt{m_4} \sqrt{m_1},\nn
\tilde{p}_8&:=& 2 \sqrt{n_1} n_2 \sqrt{n_3} \sqrt{m_3}
\sqrt{m_4}+2 m_2 \sqrt{n_3} \sqrt{m_3} \sqrt{m_4} \sqrt{m_1}+2
\sqrt{n_2} \sqrt{m_2} n_3 \sqrt{n_4} \sqrt{m_1} \nn && +2
\sqrt{n_1} \sqrt{n_2} \sqrt{m_2} n_3 \sqrt{n_4}+2 m_2 \sqrt{n_3}
\sqrt{m_3} \sqrt{n_4} \sqrt{m_1} +4 \sqrt{n_2} \sqrt{m_2}
\sqrt{n_3} \sqrt{m_3} \sqrt{n_4} \sqrt{m_1} \nn && +2 \sqrt{n_2}
\sqrt{m_2} n_3 \sqrt{m_4} \sqrt{m_1}+2 \sqrt{n_1} \sqrt{n_2}
\sqrt{m_2} m_3 \sqrt{m_4}+4 \sqrt{n_1} \sqrt{n_2} \sqrt{m_2}
\sqrt{n_3} \sqrt{m_3} \sqrt{m_4} \nn && +2 n_2 \sqrt{n_3}
\sqrt{m_3} \sqrt{n_4} \sqrt{m_1}+2 \sqrt{n_1} n_2 \sqrt{n_3}
\sqrt{m_3} \sqrt{n_4}+2 \sqrt{n_1} m_2 \sqrt{n_3} \sqrt{m_3}
\sqrt{n_4} \nn && +2 \sqrt{n_1} \sqrt{n_2} \sqrt{m_2} m_3
\sqrt{n_4}+2 \sqrt{n_2} \sqrt{m_2} m_3 \sqrt{n_4} \sqrt{m_1}+2
\sqrt{n_2} \sqrt{m_2} m_3 \sqrt{m_4} \sqrt{m_1} \nn && +2
\sqrt{n_1} \sqrt{n_2} \sqrt{m_2} n_3 \sqrt{m_4}+4 \sqrt{n_1}
\sqrt{n_2} \sqrt{m_2} \sqrt{n_3} \sqrt{m_3} \sqrt{n_4}+2
\sqrt{n_1} m_2 \sqrt{n_3} \sqrt{m_3} \sqrt{m_4} \nn && +\sqrt{n_1}
n_2 n_3 \sqrt{m_4}+\sqrt{n_1} m_2 m_3 \sqrt{n_4}+n_2 m_3
\sqrt{m_4} \sqrt{m_1} +m_2 n_3 \sqrt{n_4} \sqrt{m_1}  \nn &&
+\sqrt{n_1} m_2 n_3 \sqrt{n_4} +\sqrt{n_1} m_2 m_3 \sqrt{m_4}+m_2
m_3 \sqrt{n_4} \sqrt{m_1}+n_2 m_3 \sqrt{n_4} \sqrt{m_1}  \nn &&
+m_2 n_3 \sqrt{m_4} \sqrt{m_1} +m_2 m_3 \sqrt{m_4}
\sqrt{m_1}+\sqrt{n_1} n_2 m_3 \sqrt{m_4}+\sqrt{n_1} n_2 m_3
\sqrt{n_4} \nn && +\sqrt{n_1} n_2 n_3 \sqrt{n_4}+\sqrt{n_1} m_2
n_3 \sqrt{m_4}+n_2 n_3 \sqrt{n_4} \sqrt{m_1}+n_2 n_3 \sqrt{m_4}
\sqrt{m_1} \nn && +2 n_2 \sqrt{n_3} \sqrt{m_3} \sqrt{m_4}
\sqrt{m_1}+4 \sqrt{n_2} \sqrt{m_2} \sqrt{n_3} \sqrt{m_3}
\sqrt{m_4} \sqrt{m_1}.
\end{eqnarray}

\end{document}